\documentclass[
 reprint,
showpacs,
 aps,prl,
 groupedaddress,
]{revtex4-1}

\usepackage{graphicx}
\usepackage{amsmath,amstext,amssymb,amsthm}
\usepackage{siunitx}
\usepackage[hidelinks]{hyperref}

\begin{document}

\title{Decoding the ultrafast formation of a Fermi-Dirac distributed electron gas
		}

\author{G. Rohde}
\email{rohde@physik.uni-kiel.de}

\author{A. Stange}

\author{A. M\"uller}

\author{M. Behrendt}

\author{L.-P. Oloff}

\author{K. Hanff}

\author{T. Albert}

\author{P. Hein}

\author{K. Rossnagel}

\author{M. Bauer}
 \homepage{http://www.physik.uni-kiel.de/en/institutes/bauer-group}
\affiliation{
 Institut f\"ur Experimentelle und Angewandte Physik, Christian-Albrechts-Universit\"at zu Kiel, 24098 Kiel, Germany
}

\date{\today}

\begin{abstract}
Time- and angle-resolved photoelectron spectroscopy with \SI{13}{fs} temporal resolution is used to follow the different stages in the formation of a Fermi-Dirac distributed electron gas in graphite after absorption of an intense \SI{7}{fs} laser pulse. Within the first \SI{50}{fs} after excitation a sequence of time frames is resolved which are characterized by different energy and momentum exchange processes among the involved photonic, electronic, and phononic degrees of freedom. The results reveal experimentally the complexity of the transition from a nascent non-thermal towards a thermal electron distribution due to the different timescales associated with the involved interaction processes.
\end{abstract}



\maketitle

\begin{figure}
		\includegraphics[width=1\linewidth]{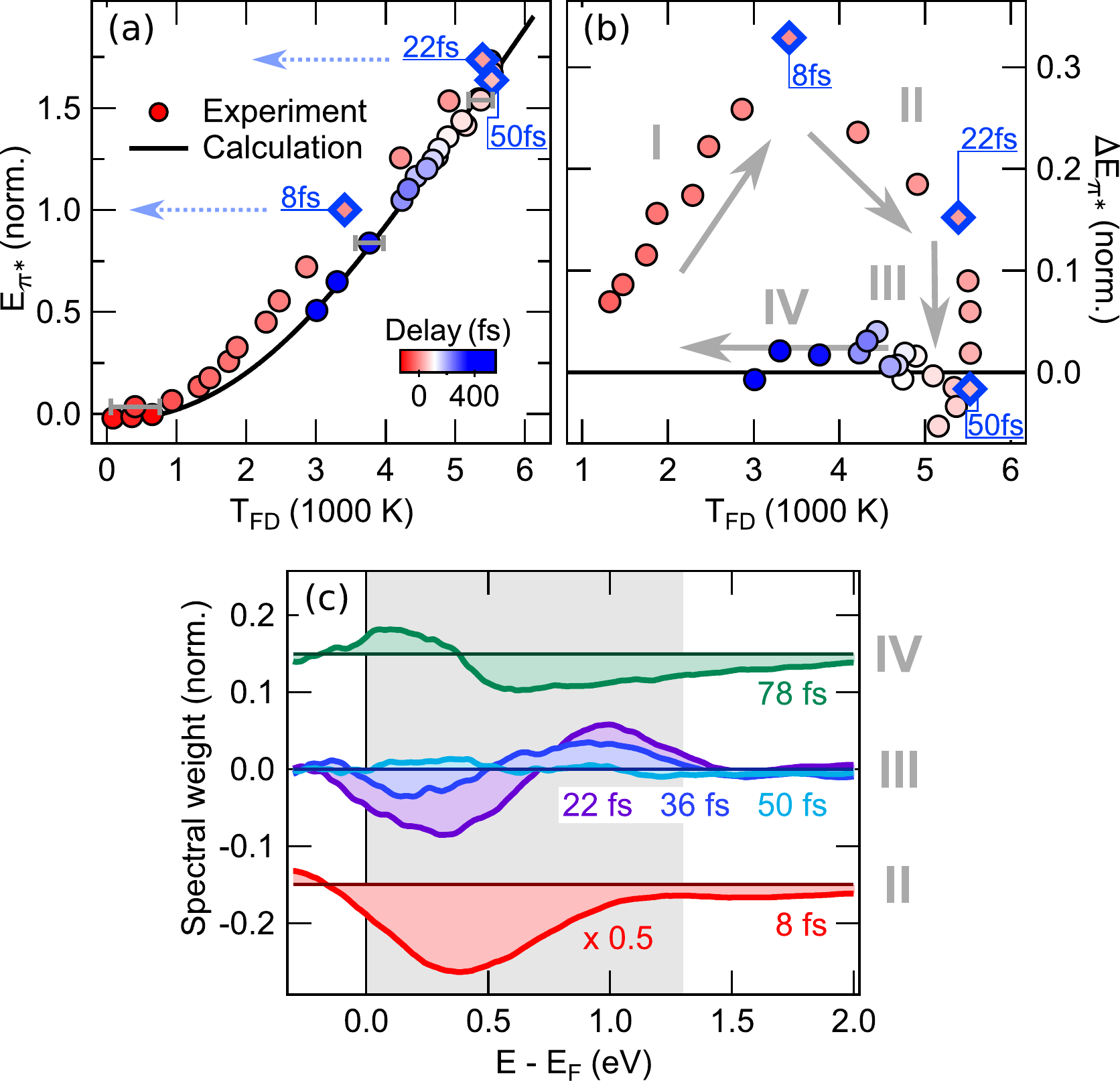}
	\caption{Analysis of trARPES data for $F=\SI{1.7}{mJ/cm^2}$. (a) $E_{\pi^*}$ as a function of $T_{\text{FD}}$. Time markers (blue diamonds) separate the different stages of thermalization identified in (b). Data are normalized to $E_{\pi^*}$ for $\Delta t = \SI{8}{fs}$. Error bars indicate exemplary uncertainties in $T_{\text{FD}}$. The solid line is a result of a calculation for a FD distributed electron gas. (b) $\Delta E_{\pi^*}$ as a function of $T_{\text{FD}}$. Arrows indicate the direction of time evolution and different stages in the thermalization process. Similar results for $F=\SI{0.9}{mJ/cm^2}$ are compiled in the Supplemental Material \cite{Rohde2018_supp}. (c) Spectrally resolved evolution of the $\pi^*$ band occupations before, during, and after stage III: difference between experimental EDCs and a FD distributed electron gas for $T=\SI{5500}{K}$. Data sets and zero lines are partly offset for clarity. Data for $\Delta t = \SI{8}{fs}$ are scaled by a factor of $0.5$. Experimental EDCs including fits are for reference added to the Supplemental Material \cite{Rohde2018_supp}.
	}\label{fig:stages25}
\end{figure}

The extraordinary nonlinearities and optical response times of graphitic materials suggest useful applications in photonics and electronics 
including light harvesting \cite{Winzer2012a,Tielrooij2015}, ultrafast photodetection \cite{Xia2009b,Gan2013}, THz lasing \cite{Ryzhii2009,Karasawa2011}, and saturable absorption \cite{Sun2009,Sun2010}.
Both characteristics are closely linked to the ultrafast dynamics of photoexcited carriers which for this material class is governed by weakly screened carrier-carrier scattering and carrier-phonon interaction. Fundamental aspects related to these processes were addressed in different time-domain studies in the past \cite{Seibert1990,Strait2011,Winnerl2013,Johannsen2013,Stange2015}. Because of limitations in the time resolution, most of these studies were restricted, however, to the characteristic timescales of electron-lattice equilibration, i.e., timescales ranging from \SI{\approx 100}{fs} to \SI{\approx 10}{ps}.   \\
The primary processes directly after photoexcitation
are, in contrast, still largely unexplored and were investigated experimentally only in a few studies so far \cite{Breusing2009,Brida2013,Gierz2015}. The dynamics in this strongly non-thermal regime is determined by phenomena such as transient population inversion, carrier multiplication, Auger recombination, but also phonon-mediated carrier redistribution \cite{Winzer2010,Ploetzing2014,Li2012,Gierz2015a}.
The challenge is to decode the relative importance and temporal sequence of these processes that drive the electronic system from a nascent non-thermal distribution as generated by photoexcitation towards a Fermi-Dirac (FD) distribution within only $\approx\SI{50}{fs}$ \cite{Breusing2009, Brida2013}.
It is obvious that such investigations rely on experiments capable of sampling this time window at an adequate time resolution of the order of \SI{10}{fs}, as well as high energy and momentum resolution.\\
This letter reports on the non-thermal carrier dynamics in highly-oriented pyrolytic graphite (HOPG) as probed in a time- and angle-resolved photoemission spectroscopy (trARPES) experiment that is operated near the transform limit at a resolution of \SI{13}{fs} (FWHM of the pump-probe cross correlation) \cite{Rohde2016}. Over the first \SI{100}{fs}, we monitor the different stages in the temporal evolution of an initially non-thermal carrier distribution generated by the absorption of a \SI{7}{fs} near-infrared pulse. We are able to dissect the non-thermal to thermal transition into sub-stages that are characterized by different energy and momentum exchange and redistribution processes among the involved photonic, electronic, and phononic degrees of freedom. Specifically, the experimental data reveal that the initiating photo-absorption process is first followed by momentum redistribution of the excited carriers before effective cooling of the non-thermal carrier distribution due to emission of phonons sets in. An internally thermalized hot electron gas is finally formed after $\approx\SI{50}{fs}$ which on much longer timescales equilibrates with the lattice \cite{Kampfrath2005,Wang2010}. The results show experimentally that on the extremely short non-thermal time frame of a few \SI{10}{fs} carrier thermalization can be a complex multi-step process owing to different timescales and efficiencies associated with momentum and energy relaxation.    \\
\begin{figure}
		\includegraphics[width=1\linewidth]{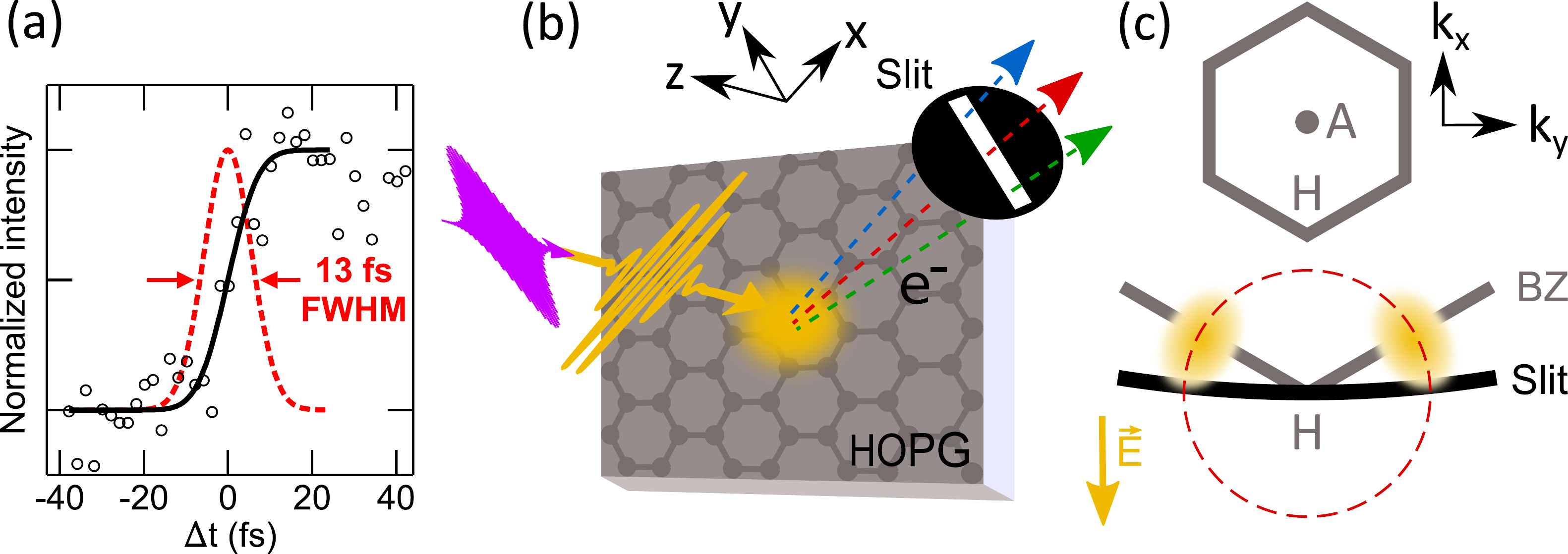}
	\caption{(a) WL-pump/XUV-probe cross correlation signal of the experiment. For details see Refs. \cite{Rohde2016} and \cite{Rohde2018_supp}. (b) Schematic of time-resolved pump-probe ARPES of HOPG. Pump pulses are polarized along the $x$ direction. The entrance slit of the electron analyzer is aligned along the $y$ direction. 
	(c) Top: illustration of the Brillouin zone of HOPG. Bottom: close-up of the Brillouin zone around the $H$ point (gray) with the momentum cut probed in the experiment indicated (black). The red dashed line marks the constant-energy contour of the $\pi^*$ band near $E-E_F=\SI{0.8}{eV}$. Areas of primary excitation are highlighted in yellow. }
	\label{fig:Scheme}
\end{figure}
Bulk samples of HOPG (Goodfellow Ltd.) were mechanically cleaved under high vacuum (\SI{1E-7}{mbar}) right before the experiments.
TrARPES was performed under ultra-high vacuum conditions (\SI{3E-10}{mbar}) in a pump-probe configuration [Fig. \ref{fig:Scheme}(b)] using \SI{7}{fs} white light (WL) pump pulses (\SI{800}{nm} center wavelength, \SI{232}{nm} RMS spectral width) and \SI{11}{fs} extreme ultraviolet (XUV) probe pulses (\SI{22.1}{eV}), both generated by the output of a 8 kHz Ti:sapphire multipass amplifier.  Photoelectron spectra were recorded with a hemispherical electron analyzer at an energy resolution of \SI{240}{meV}. The relative orientation of the HOPG sample and analyzer entrance slit chosen for the experiments resulted in a momentum cut centered at the $H$ point as indicated by the black line in Fig. \ref{fig:Scheme}(c). Experiments were performed at near-normal incidence at two different incident pump fluences of $F=\SI{0.9}{mJ/cm^2}$ and $F=\SI{1.7}{mJ/cm^2}$. The pump pulse polarization was oriented perpendicular to the momentum cut probed in photoemission as illustrated in Fig. \ref{fig:Scheme}(c). The FWHM of the cross correlation signal measured at the sample position yields a value of \SI{13}{fs} [see Fig. \ref{fig:Scheme}(a)] corresponding to a time resolution of \SI{8}{fs} (\SI{5.5}{fs}) based on a definition introduced in Ref. \cite{Gierz2015} (Ref. \cite{Gierz2017}). All data were recorded at an equilibrium sample temperature of \SI{300}{K}. Further details of the experimental setup are described in Ref. \cite{Rohde2016}.\\
\begin{figure}
        \includegraphics[width=0.8\linewidth]{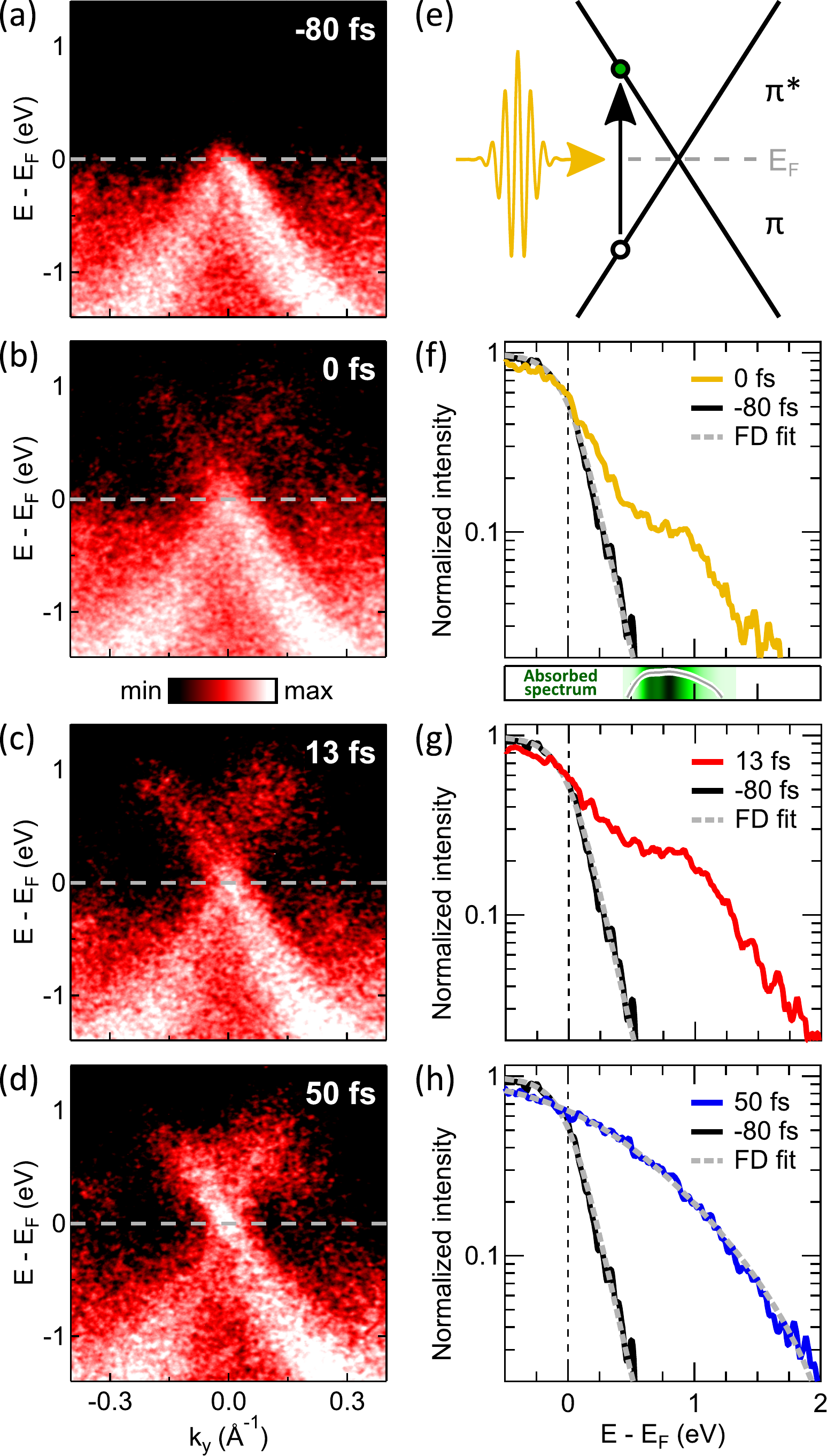}
	\caption{Time-resolved ARPES data of HOPG taken near $H$ for excitation with \SI{7}{fs} pump pulses ($F=\SI{1.7}{mJ/cm^2}$). 
	(a)-(d) ARPES snapshots recorded at different time delays $\Delta t$. (e) Illustration of the photoexcitation process. 
	(f)-(h) EDCs around $E_F$ for different $\Delta t$ derived from the trARPES data. Photoemission intensity was integrated over a momentum window of $\SI{0.8}{\angstrom^{-1}}$. Dashed lines are fits of a FD distribution to the EDCs. The inset underneath (f) indicates the pump pulse spectrum (logarithmic scale) convolved with the energy resolution of the trARPES experiment and mapped onto the energy axis according to the excitation process illustrated in (e). Momentum-resolved photoemission intensity transients are for reference added to the Supplemental Material \cite{Rohde2018_supp}. 
	}
	\label{fig:snapshots+edcs25}
\end{figure}
Figure \ref{fig:snapshots+edcs25}(a) shows trARPES data of HOPG in the vicinity of the $H$ point and around the Fermi energy $E_F$ with the pump-probe time delay set to $\Delta t=\SI{-80}{fs}$, i.e., prior to the excitation by the \SI{7}{fs} white light pump pulse. The occupied and downward-dispersing $\pi$ band of HOPG, which at $H$ exhibits an almost linear dispersion \cite{Zhou2006a,Grueneis2008}, is well resolved. No indications of the upward-dispersing $\pi^*$ band are visible, as expected for an undoped sample at thermal equilibrium. Figures \ref{fig:snapshots+edcs25}(b)-\ref{fig:snapshots+edcs25}(d) show trARPES data recorded at a pump fluence of \SI{1.7}{mJ/cm^2} for selected time delays $\SI{0}{fs} \leq \Delta t \leq \SI{50}{fs}$. Distinct changes in the excited state electron distribution taking place on a sub-\SI{10}{fs} timescale are clearly resolved in the experimental data. They become even more evident in a semi-logarithmic energy distribution curve (EDC) representation obtained by momentum integration of the raw data following a data analysis scheme described in Ref. \cite{Stange2015}. Figures \ref{fig:snapshots+edcs25}(f)-\ref{fig:snapshots+edcs25}(h) display the EDC intensities $I$ as a function of energy $E$ for the different time delays after excitation in comparison to the equilibrium state distribution at $T=\SI{300}{K}$ ($\Delta t=\SI{-80}{fs}$). The initial equilibrium state EDC as well as the EDC for $\Delta t=\SI{50}{fs}$ are well described by a FD distribution (see gray dashed lines). In contrast, the data recorded at time zero and at $\Delta t=\SI{13}{fs}$ show clear deviations from a FD distributed population indicating the strong non-thermal character of the electronic system on sub-\SI{50}{fs} timescales \cite{Fann1992}.   \\
To evaluate the thermal character of the electron gas at a given time delay two different quantities are separately extracted from each measured EDC: Firstly, we determine the best-fit Fermi-Dirac temperature $T_{\text{FD}}$ \cite{Fann1992a} from a fit of a FD distribution to the data. Secondly, we numerically compute the integral 
\begin{equation}
E_{\pi^*}=\int_{E>E_F}I(E)\cdot E\, \mathrm{d}E    
\end{equation}
being a measure for the total energy stored in the $\pi^*$ band along the probed momentum cut. Figure \ref{fig:stages25}(a) depicts the relation between $E_{\pi^*}$ and $T_{\text{FD}}$ as it evolves in time in comparison to what is expected for a FD distributed electron gas (solid line). The color coding of the experimental data indicates the time delay $\Delta t$ of the measurement with red corresponding to small delays and blue corresponding to long delays. We consider the measured electron distribution being internally thermalized if the data points lie on top of the result for a FD distribution. The representation of the experimental data in Fig. \ref{fig:stages25}(a) matches the expectation for the build-up and decay of a non-thermal electron distribution following an intense photoexcitation: In response to the absorption of the pump pulse, $E_{\pi^*}$ starts to increase around time zero and at the same time the data points leave the thermalized regime. Eventually the gain in $E_{\pi^*}$ stops and finally, at an equilibrium temperature of $\approx \SI{5500}{K}$, the experimental data points merge into the thermalized regime again. For longer timescales, the electronic system stays internally thermalized and cools down due to coupling to the lattice. The different stages of the non-thermal to thermal transition become more evident in the representation of the data shown in Fig. \ref{fig:stages25}(b). The graph displays the difference $\Delta E_{\pi^*}$ between experimental results and FD distribution as a function of $T_{\text{FD}}$. $\Delta E_{\pi^*}$ may be considered as a measure of the non-thermal component of $E_{\pi^*}$. The distinct changes observable in the slope of $\Delta E_{\pi^*}$ vs. $T_{\text{FD}}$ suggest the discrimination of four different stages of the thermalization process [labeled I-IV in Fig. \ref{fig:stages25}(b)]. Time markers separating these stages (blue diamonds) are for comparison also added to the data in Fig. \ref{fig:stages25}(a).

The first stage of the probed thermalization process (I) reflects the primary energy input into the electronic system directly resulting from the absorption of the pump pulse. Up to a time delay of $\Delta t \approx\SI{8}{fs}$ we observe a continuous increase of $\Delta E_{\pi^*}$, i.e., a continuous build-up of the non-thermal character of the electron gas. This timescale agrees well with the result of an evaluation for the temporal evolution of the absorbed pump energy under consideration of the experimental parameters: At $\Delta t=\SI{8}{fs}$ the experiment probes a state of the electronic system  where already \SI{93}{\%} of the total pump excitation energy has been absorbed.

The second stage of the thermalization process (II) lasts until $\Delta t \approx \SI{22}{fs}$ and is characterized by a continuous decrease of $\Delta E_{\pi^*}$ [see Fig. \ref{fig:stages25}(b)] while $T_{\text{FD}}$ as well as $E_{\pi^*}$ still increase [see Fig. \ref{fig:stages25}(a)]. In the following, we will argue how these observations are related to electron-electron and electron-phonon scattering processes, respectively. The overall detected energy input within stage II corresponds to $\approx\SI{75}{\%}$ of the energy that has been accumulated during the direct absorption process within the first stage [cf. arrows in Fig. \ref{fig:stages25}(a)]. On these intermediate timescales, the photoexcited electron system itself is the only reservoir which can account for a substantial energy gain in the probed region of energy-momentum space. It is result of a carrier redistribution process that is driven by the primary momentum anisotropy generated in graphite upon photoexcitation with linearly polarized light \cite{Gruneis2003, Malic2011}. For the pump pulse polarization used in the present study the situation is schematically illustrated in Fig. \ref{fig:Scheme}(c). Primary population maxima are generated at the intersection of the $\pi^*$-cone and Brillouin zone boundary (yellow marked areas) and outside of the momentum cut probed in the experiment \cite{Gruneis2003}. The observed secondary gain in $E_{\pi^*}$ results from a net population transfer out of these primary areas by scattering processes involving a finite azimuthal momentum transfer component, i.e., noncollinear scattering processes \cite{Mittendorff2014}.
On the relevant timescales and for the pump fluences used in the experiment, noncollinear scattering processes are predominantly due to the interaction of the electrons with strongly coupled optical phonons (SCOPs) \cite{Malic2012, Trushin2015, Aeschlimann2017}. 
The time frame for the decay of the pump-induced momentum anisotropy as implied by our data matches well findings for graphene: Results of calculations show that for pump fluences in the $\SI{1}{mJ/cm^2}$ range an isotropic momentum distribution is formed well within the first \SI{50}{fs} after excitation \cite{Aeschlimann2017}. Polarization-dependent photoluminescence measurements performed at excitation fluences of $\approx \SI{100}{\micro J/cm^2}$ hint to a characteristic timescale for this process of $\approx \SI{12}{fs}$ \cite{Danz2017}. 

The continuous decrease of the non-thermal character of the electron gas during the second stage as implied by the decrease of $\Delta E_{\pi^*}$ indicates that simultaneously with the phonon-driven decay of the momentum anisotropy the energy distribution of the electrons starts to converge towards a FD distribution. In the pump fluence regime of our experiment, this process is dominated by collinear scattering processes due to Coulomb interaction among the carriers \cite{Winzer2013a}. This will further persist during the third stage of the thermalization process and finally results in the formation of a hot FD distribution at $\Delta t \approx \SI{50}{fs}$.

\begin{figure}
		\includegraphics[width=1\linewidth]{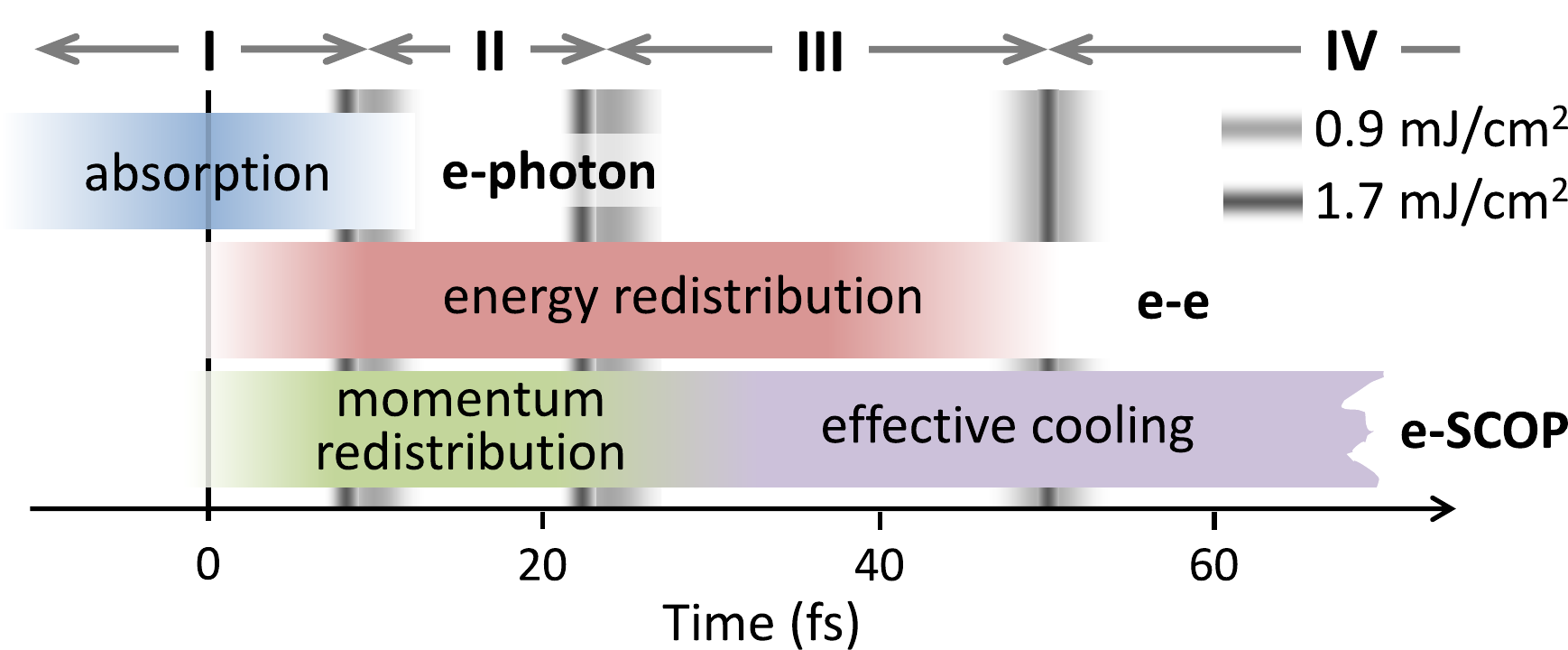}
	\caption{Stages of thermalization and characteristic timescales of energy and momentum exchange processes. Relevant interaction processes are indicated. Time markers separating the stages result from the evaluation of the experimental data for $F=\SI{0.9}{mJ/cm^2}$ and $F=\SI{1.7}{mJ/cm^2}$. Widths of the markers account for the experimental uncertainty.
	}
	\label{fig:summary}
\end{figure}
 
The signature discriminating the third stage (III) from the other stages is the onset of a net energy drain out of the electron system [decrease of $E_{\pi*}$, see Fig. \ref{fig:stages25}(a)], while the electron distribution is still non-thermal (``non-thermal cooling''). It lasts for $\approx \SI{28}{fs}$ until at $\Delta t= \SI{50}{fs}$ the comparison of the experimental data with the calculated results for a FD distribution indicate that finally the electronic system is internally fully thermalized. The energy drain is caused by electron-lattice interaction which for timescales $\lessapprox \SI{500}{fs}$ is dominated by the interaction with SCOPs \cite{Kampfrath2005,Wang2010}. Notably, during this initial energy relaxation stage the energy drain seems to affect only $\Delta E_{\pi^*}$, which we associated with the non-thermal component of the electronic excitation, while $T_{\text{FD}}$ stays virtually constant at its maximum value of $\approx \SI{5500}{K}$. Figure \ref{fig:stages25}(c) illustrates how this cooling process acts on the electron distribution. The graph displays difference intensities of experimental EDCs and a FD distribution at $T=\SI{5500}{K}$ for different time delays before ($\Delta t= \SI{8}{fs}$), during ($\Delta t= \SI{22}{fs}, \SI{36}{fs}, \SI{50}{fs}$), and after ($\Delta t= \SI{78}{fs}$) the non-thermal cooling stage. Whereas the experimental EDCs at $\Delta t= \SI{8}{fs}$ ($T_{\text{FD}}=$ \SI{3400}{K}) and $\Delta t= \SI{78}{fs}$ ($T_{\text{FD}}=$ \SI{5200}{K}) show significant deviations from the FD distribution at $T=\SI{5500}{K}$ over the entire excitation energy regime, the deviations for $\Delta t= \SI{22}{fs}$ and $\Delta t= \SI{36}{fs}$ ($T_{\text{FD}}\approx$ \SI{5500}{K}) are limited to an intermediate energy regime between $E_F$ and $E-E_F\approx \SI{1.3}{eV}$. The high energy tails of these two EDCs match in contrast the thermalized behavior extremely well. In this part of the electron spectrum thermal equilibrium conditions corresponding to the maximum electron temperature of $\SI{5500}{K}$ are anticipated already \SI{28}{fs} before the complete internal thermalization is achieved. 

The energy drain by the interaction of the electrons with SCOPs during stage III should affect both, the `non-thermal' as well as the `thermal' part of the electron distribution. The internal thermalization of the electron gas promoted by carrier-carrier interaction results, on the other hand, in a net energy transfer from the `non-thermal' to the `thermal' part of the electron distribution. The observation of a rigid and thermalized high energy tail during the entire non-thermal cooling stage indicates that the latter process effectively compensates for the energy drain out of the thermal part due to interaction with the lattice.

The final stage of the thermalization process that can be discerned in Fig. \ref{fig:stages25} (IV, $\Delta t \geq \SI{50}{fs}$) is characterized by the continuous and simoultaneous decrease of $E_{\pi^*}$ and $T_{\text{FD}}$ with the experimental data following the expectations for a FD-distributed electron gas strikingly well. This is indicative for the cool-down of the internally thermalized electron distribution (thermal cooling) due to the interaction with the lattice. This temporal regime was previously studied in detail by various time-domain techniques \cite{Seibert1990,Kampfrath2005,Strait2011,Winnerl2013,Johannsen2013,Stange2015}.  

The transition of a photoexcited electron gas in HOPG from a nascent non-thermal towards a thermal distribution is completed within a time period as short as \SI{50}{fs}. The scenario involves electron-photon, electron-electron, and electron-phonon interaction affecting the response of the system on different timescales. In analyzing the temporal evolution of the excess energy deposited into the electronic system within $\SI{7}{fs}$, we were able to experimentally identify characteristic signatures of these interaction processes. Owing to the exceptional time resolution of the trARPES experiment used in the study we even succeeded in dissecting the complex thermalization process into different stages. The sequence of these stages and the characteristic timescales of energy and momentum exchange among the involved systems are summarized in Fig. \ref{fig:summary}. For the two fluences investigated in this work, no significant differences in the characteristic response times could be observed (see Fig. \ref{fig:summary} and Supplemental Material \cite{Rohde2018_supp}). However, we expect significant changes at substantially lower fluences particularly as carrier-carrier scattering is very sensitive to the excited carrier density \cite{Winzer2013}. On the other hand,  indirect electronic transitions may become relevant at higher fluences due to saturation effects and Pauli blocking \cite{Yang2017}. Moreover, the thermalization of the system can be affected by other experimental parameters such as bandwidth and photon energy of the applied pump pulses \cite{Koenig-Otto2016}.
Similarly complex multi-step carrier thermalization scenarios are expected also for other materials with reduced screening. The response of these systems to an optical excitation is phenomenologically often described within multi-temperature models \cite{Anisimov1974} which consider the electronic system being internally thermalized for all times and therefore neglect for instance phonon emission during an internal thermalization stage \cite{Baranov2014}. The method presented here can provide valuable information on the actual state of the electron gas even on the extremely short `non-thermal' timescales right after excitation.

This work was supported by the German Research Foundation (DFG) through project BA 2177/10-1.

\bibliography{library_ed}

\end{document}


\title{Supplemental Material for\texorpdfstring{\\}{ }``Decoding the ultrafast formation of a Fermi-Dirac distributed electron gas''}

\author{G. Rohde}
\email{rohde@physik.uni-kiel.de}

\author{A. Stange}

\author{A. M\"uller}

\author{M. Behrendt}

\author{L.-P. Oloff}

\author{K. Hanff}

\author{T. Albert}

\author{P. Hein}

\author{K. Rossnagel}

\author{M. Bauer}
 \homepage{http://www.physik.uni-kiel.de/en/institutes/bauer-group}
\affiliation{
 Institut f\"ur Experimentelle und Angewandte Physik, Christian-Albrechts-Universit\"at zu Kiel, 24098 Kiel, Germany
}

\maketitle

\section{Time resolution of the trARPES experiment}

The cross correlation between white light (WL) pump and extreme ultraviolet (XUV) probe was determined from trARPES measurements of a 1\textit{T}-TiSe$_2$ reference sample. In a previous work we showed that the excited state population rise time associated with an optical interband transition in this material provides a very good approximation for the time resolution of the experiment \cite{Rohde2016}. Fig. 1(a) of the main text shows a photoemission signal transient probing the excited state population generated by this transition upon WL photoexcitation. In order to determine the time resolution of the experiment from this data, we fitted an error function to the signal rise [solid line in Fig. 1(a)]. The derivative of the error function is a Gaussian [dashed line in Fig. 1(a)], which we interpret as an approximation of the WL-XUV cross correlation trace. The width of the Gaussian is \SI{13}{fs}. For more details on the cross correlation measurement we refer to Ref. \cite{Rohde2016}.

\begin{figure}
		\includegraphics[width=.55\linewidth]{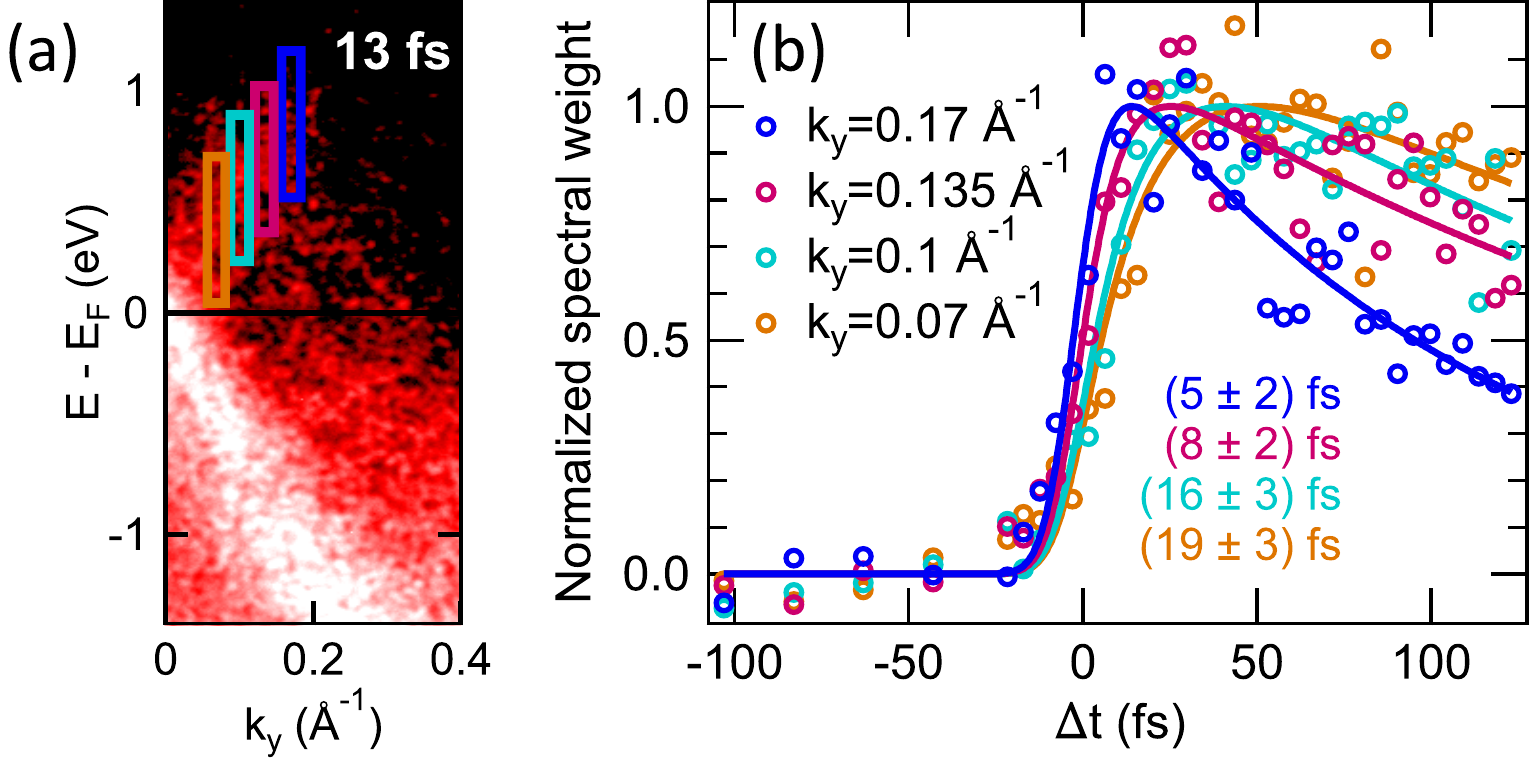}
	\caption{Momentum-resolved population dynamics along the $\pi^*$ band of HOPG for excitation with \SI{7}{fs} WL pulses ($F=\SI{0.9}{mJ/cm^2}$). (a) Photoemission snapshot near $H$ recorded at $\Delta t =\SI{13}{fs}$. (b) Photoemission intensity transients for selected momentum areas marked in (a). Solid lines are fits to the experimental data. The exponential rise times resulting from the fits are indicated.}
	\label{fig:transients26}
\end{figure}

To illustrate the exceptional time resolution of the experimental setup we additionally compare in Fig. \ref{fig:transients26}(b) photoemission intensity transients of HOPG for selected energy-momentum areas as indicated by the rectangles in the photoemission intensity map in Fig. \ref{fig:transients26}(a). The solid lines in Fig. \ref{fig:transients26}(b) are fits to the experimental data. The fitting function accounts for the signal rise and decay by two exponential functions, which were convolved with a Gaussian representing the experimental time resolution. The intensity rise times resulting from the fits are indicated and show that also small differences in the population dynamics of a few femtoseconds are well resolved. Independent of momentum we always observe a delayed signal rise with respect to the measured cross correlation signal resulting from the momentum redistribution of electrons into the probed momentum cut by e-SCOP interaction processes, which become dominant in stage II of the thermalization process.

\section{Energy distribution curves: supplemental information to Fig. 3 of the main text}

\begin{figure}[htbp]
\includegraphics[width=.67\linewidth]{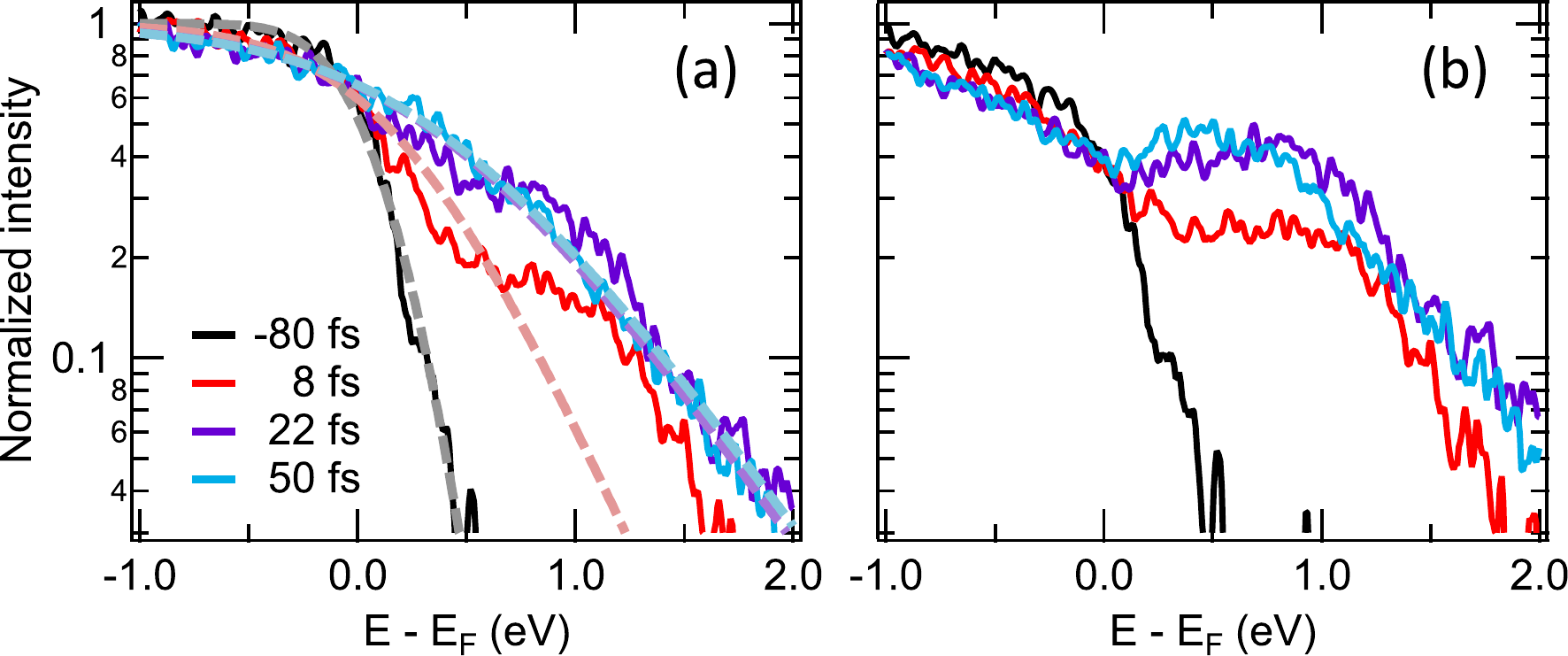}
\caption{(a) Spectrally resolved evolution of electron occupations around $E_F$ obtained from trARPES data at $F=\SI{1.7}{mJ/cm^2}$: EDCs (solid lines) for different $\Delta t$ in comparison to FD fits (dashed lines). (b) Corresponding raw data EDCs not processed according to Ref. \cite{Stange2015}.}
	\label{fig:EDCswFits25}
\end{figure}

Figure \ref{fig:EDCswFits25}(a) shows EDCs of trARPES data that were used to calculate the difference intensity traces shown in Fig. 3(c) of the main text. For comparison we also added the results of the fits of a FD distribution to the experimental data to the graph. Figure \ref{fig:EDCswFits25}(b) shows corresponding raw data EDCs that were not processed by the data analysis scheme described in detail in Ref. \cite{Stange2015}. For the quantitative analysis of the data this procedure is indispensable, particularly as matrix elements for photoemission from the $\pi$ band and the $\pi^*$ band significantly differ \cite{Gruneis2008a}.

\section{time-resolved ARPES data for \texorpdfstring{$\bm{F}\mathbf{\boldsymbol{=}0.9\, mJ\boldsymbol{/}cm^2}$}{F=0.9 mJ/cm2}}

\begin{figure}[htbp]
		\includegraphics[width=.85\linewidth]{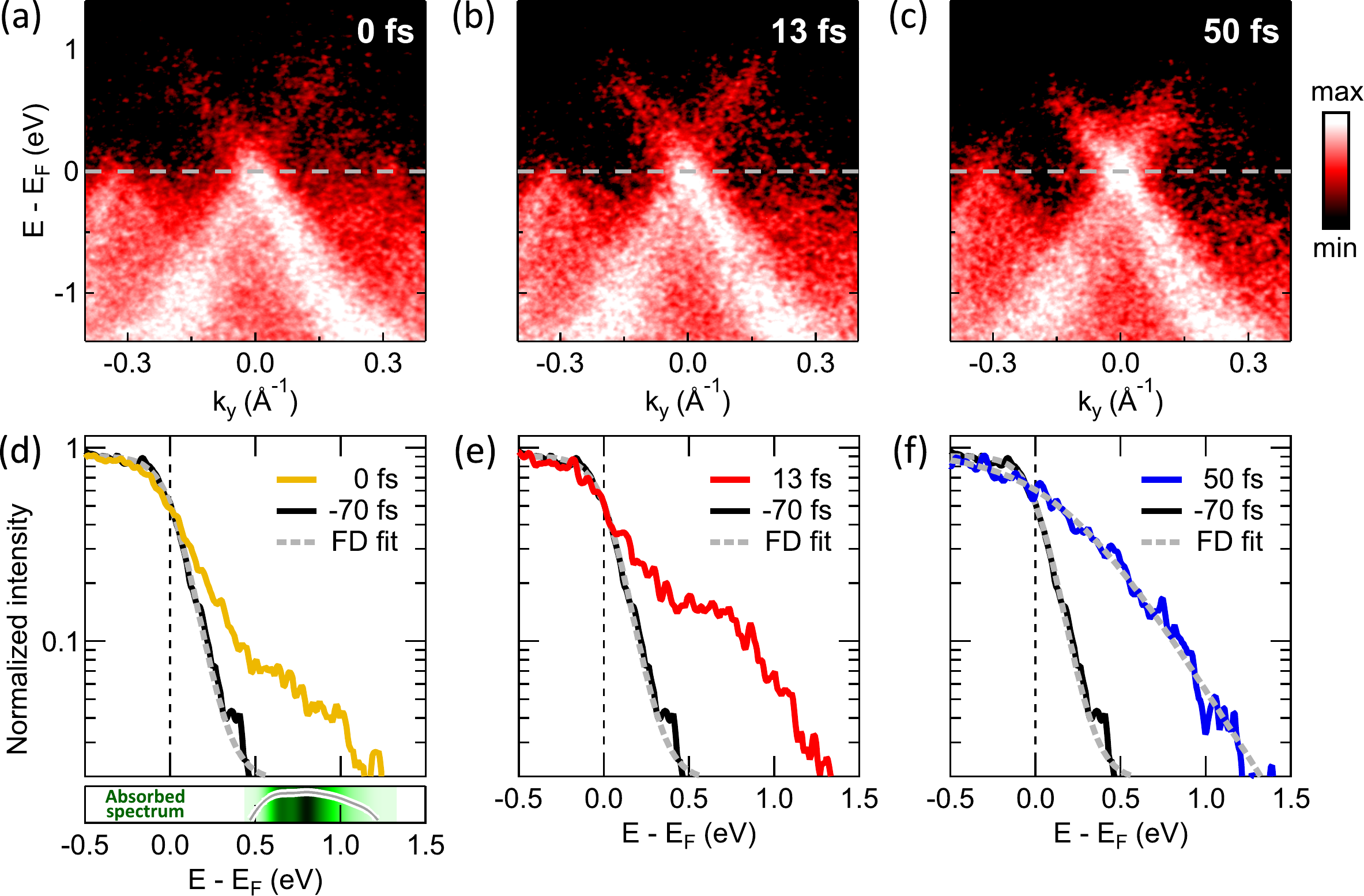}
	\caption{Time-resolved ARPES data of HOPG near $H$ for excitation with \SI{7}{fs} pump pulses ($F=\SI{0.9}{mJ/cm^2}$).
	(a)-(c) ARPES snapshots recorded at different time delays $\Delta t$.
	(d)-(f) EDCs around $E_F$ for different $\Delta t$ derived from the trARPES data. Photoemission intensity was integrated over a momentum window of $\SI{0.8}{\angstrom^{-1}}$.
	Dashed lines are fits of a FD distribution to the EDCs. The inset  underneath (d) indicates the pump pulse spectrum (logarithmic scale) convolved with the energy resolution of the trARPES experiment and mapped onto the energy axis according to the excitation process illustrated in Fig. 2(e) of the main text. }
	\label{fig:snapshots+edcs26}
\end{figure}

Experiments were performed at two different incident pump fluences $F$. In the main text data for $F=\SI{1.7}{mJ/cm^2}$ are shown. In this section data are presented that were recorded at $F=\SI{0.9}{mJ/cm^2}$  with the other experimental parameters being kept  the same. Figures \ref{fig:snapshots+edcs26}(a)-\ref{fig:snapshots+edcs26}(c) show trARPES data recorded at a pump fluence of $F=\SI{0.9}{mJ/cm^2}$ for three selected time delays $\SI{0}{fs} \leq \Delta t \leq \SI{50}{fs}$.
Figures \ref{fig:snapshots+edcs26}(d)-\ref{fig:snapshots+edcs26}(f) display the corresponding EDCs in comparison to the equilibrium state distribution at $T=\SI{300}{K}$ that was recorded at $\Delta t=\SI{-70}{fs}$.
Similar to the data shown in Fig. 2 of the main text the equilibrium state EDC as well as the EDC for $\Delta t=\SI{50}{fs}$ are well described by a FD distribution, whereas clear non-thermal signatures are observed at time zero and at $\Delta t = \SI{13}{fs}$.

\begin{figure}
		\includegraphics[width=.65\linewidth]{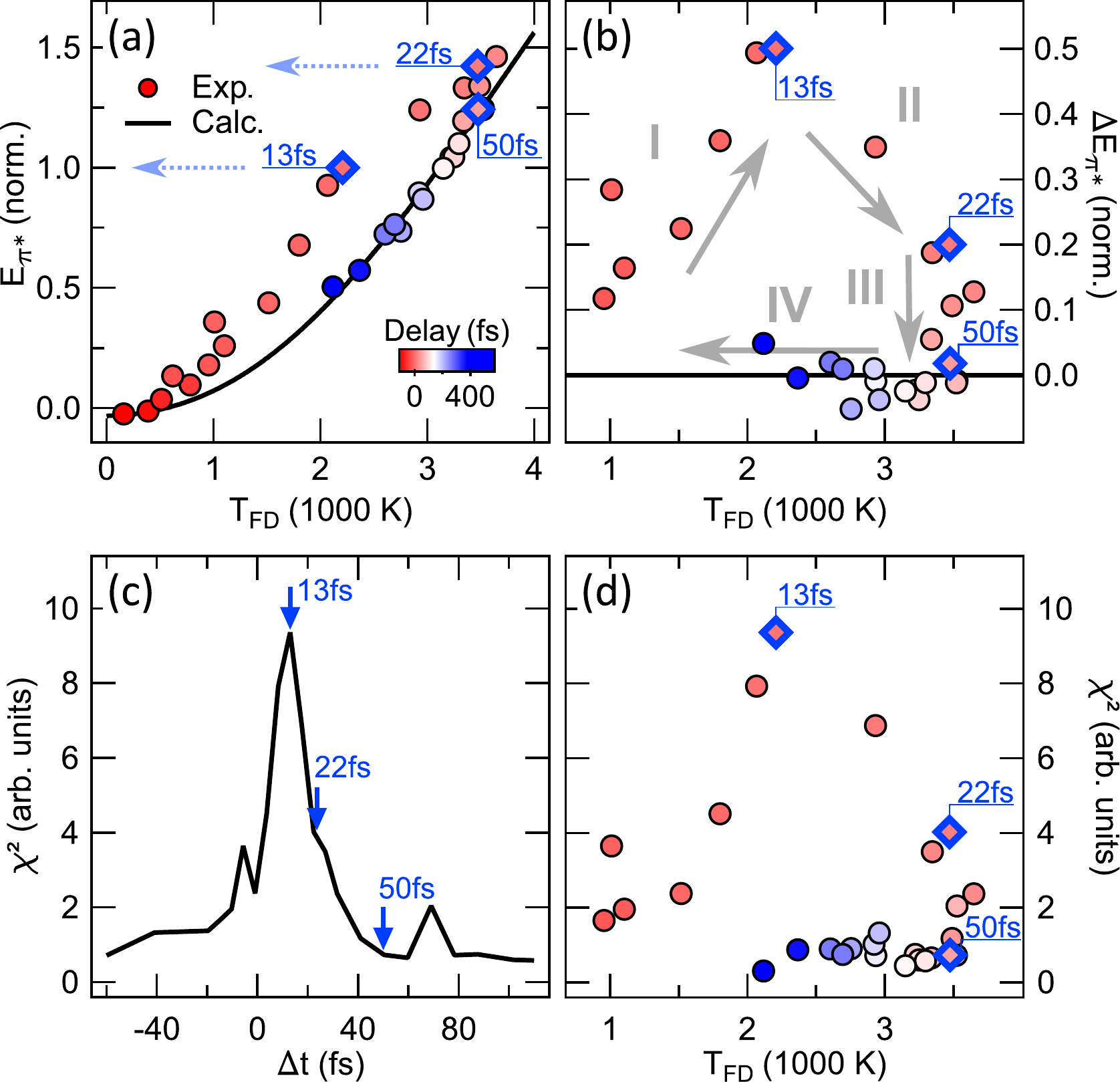}
	\caption{Analysis of trARPES data for $F=\SI{0.9}{mJ/cm^2}$. (a) $E_{\pi^*}$ as a function of $T_{\text{FD}}$. Time markers (blue diamonds) separate the different stages of thermalization identified in (b). Data are normalized to $E_{\pi^*}$ for $\Delta t = \SI{13}{fs}$. The solid line is result of a calculation for a FD distributed electron gas. (b) $\Delta E_{\pi^*}$ as a function of $T_{\text{FD}}$. Arrows indicate the direction of time evolution and different stages in the thermalization process. (c) Sum of the squared deviations, $\chi^2$, between experimental EDCs and fits of a FD distribution as a function of $\Delta t$. (d) $\chi^2$ as a function of $T_{\text{FD}}$.}
	\label{fig:stages26}
\end{figure}

Fig. \ref{fig:stages26}(a) depicts $E_{\pi^*}$ as computed from the experimental EDCs according to Eq. (1) in the main text as a function of the best-fit Fermi-Dirac temperature $T_{\text{FD}}$ for $F=\SI{0.9}{mJ/cm^2}$. The solid line shows for comparison the expectations for a FD distributed electron gas.  Fig. \ref{fig:stages26}(b) displays the non-thermal component of the $\pi^*$ band energy along the probed momentum cut, $\Delta E_{\pi^*}$, as a function of $T_{\text{FD}}$. Both graphs qualitatively reproduce the experimental findings for $F=\SI{1.7}{mJ/cm^2}$ presented in the main text. Minor deviations are observed for the time markers separating the different stages of the thermalization process. Due to the scatter of the data particularly in the low fluence data set the significance of these differences is not ensured. Time markers evaluated from the low fluence data are added for comparison to Fig. 4 of the main text. The overall detected energy input within stage~II at $F=\SI{0.9}{mJ/cm^2}$ corresponds to $\approx \SI{45}{\%}$ of the energy that has been accumulated during the direct absorption process within stage I [cf. arrows in Fig. \ref{fig:stages26}(a)]. This fraction is significantly smaller than the value of $\approx \SI{75}{\%}$ as observed at $F=\SI{1.7}{mJ/cm^2}$. 

The goodness of the fits of the FD distribution to the experimental data can be considered as an alternative quantity accounting for the thermal character of the electron distribution. In Fig. \ref{fig:stages26}(c) we plot the sum of the squared deviations of the fit, $\chi^2$, as a function of $\Delta t$. In this representation, the characteristic timescales of the primary photoexcitation process and the overall thermalization time are well discerned. Figure \ref{fig:stages26}(d) displays $\chi^2$ as a function of {$T_{\text{FD}}$}. This representation reproduces the overall temporal evolution of the thermalization process as deduced from an energy analysis [Fig. \ref{fig:stages26}(b)] and likewise allows for a clear discrimination of stages I - IV.
\\
\quad
\bibliography{library_ed}